\documentclass[12pt]{article}
\usepackage{amsfonts}
\usepackage{amsmath, amsthm}
\usepackage{epsfig}
\usepackage{algorithm}

\usepackage{soul}
\usepackage[normalem]{ulem}
\usepackage{amsfonts}
\usepackage{subcaption}
\usepackage{epsfig}
\usepackage{booktabs}
\usepackage{lscape}
\usepackage{multirow}
\usepackage{longtable}
\usepackage{rotating}
\usepackage{array}
\usepackage{multicol}
\usepackage{graphicx}
\usepackage{bigstrut}
\usepackage{setspace}
\usepackage{epstopdf}
\usepackage{mathrsfs}
\usepackage{amsfonts}
\usepackage{epsfig}
\usepackage{booktabs}
\usepackage{threeparttable}
\usepackage{epstopdf}
\usepackage{lscape}
\usepackage{multirow}
\usepackage{enumerate}

\usepackage{algorithmic}
\usepackage{amsmath,verbatim,color,amssymb,epsfig}
\usepackage{bm}
\usepackage{amsfonts}
\usepackage{epsfig}
\usepackage{multirow}
\usepackage{graphicx}
\usepackage{array}
\usepackage[table]{xcolor}
\definecolor{Gray}{gray}{0.80}
\usepackage{lscape}
\usepackage{cite}
\usepackage{natbib}
\usepackage[normalem]{ulem}
\usepackage[colorlinks=true,urlcolor=blue,citecolor=black,linkcolor=blue,bookmarks=true]{hyperref}
\usepackage{caption}
\begin{document}
\def\eqx"#1"{{\label{#1}}}
\def\eqn"#1"{{\ref{#1}}}

\makeatletter 
\@addtoreset{equation}{section}
\makeatother  

\def\yincomment#1{\vskip 2mm\boxit{\vskip 2mm{\color{red}\bf#1} {\color{blue}\bf --Yin\vskip 2mm}}\vskip 2mm}
\def\squarebox#1{\hbox to #1{\hfill\vbox to #1{\vfill}}}
\def\boxit#1{\vbox{\hrule\hbox{\vrule\kern6pt
          \vbox{\kern6pt#1\kern6pt}\kern6pt\vrule}\hrule}}

\newcommand{\blue}[1]{\textcolor{blue}{{#1}}}
\newcommand{\red}[1]{\textcolor{red}{{#1}}}
\def\theequation{\thesection.\arabic{equation}}
\newcommand{\ds}{\displaystyle}

\newcommand{\bJ}{\mbox{\bf J}}
\newcommand{\bF}{\mbox{\bf F}}
\newcommand{\bM}{\mbox{\bf M}}
\newcommand{\bR}{\mbox{\bf R}}
\newcommand{\bZ}{\mbox{\bf Z}}
\newcommand{\bX}{\mbox{\bf X}}
\newcommand{\bx}{\mbox{\bf x}}
\newcommand{\bQ}{\mbox{\bf Q}}
\newcommand{\bH}{\mbox{\bf H}}
\newcommand{\bh}{\mbox{\bf h}}
\newcommand{\bz}{\mbox{\bf z}}
\newcommand{\ba}{\mbox{\bf a}}
\newcommand{\bG}{\mbox{\bf G}}
\newcommand{\bB}{\mbox{\bf B}}
\newcommand{\bb}{\mbox{\bf b}}
\newcommand{\bA}{\mbox{\bf A}}
\newcommand{\bC}{\mbox{\bf C}}
\newcommand{\bI}{\mbox{\bf I}}
\newcommand{\bD}{\mbox{\bf D}}
\newcommand{\bU}{\mbox{\bf U}}
\newcommand{\bc}{\mbox{\bf c}}
\newcommand{\bd}{\mbox{\bf d}}
\newcommand{\bs}{\mbox{\bf s}}
\newcommand{\bS}{\mbox{\bf S}}
\newcommand{\bV}{\mbox{\bf V}}
\newcommand{\bv}{\mbox{\bf v}}
\newcommand{\bW}{\mbox{\bf W}}
\newcommand{\bw}{\mbox{\bf w}}
\newcommand{\bg}{\mbox{\bf g}}
\newcommand{\bu}{\mbox{\bf u}}

\newcommand{\bcU}{\boldsymbol{\cal U}}
\newcommand{\bbeta}{\boldsymbol{\beta}}
\newcommand{\bdelta}{\boldsymbol{\delta}}
\newcommand{\bDelta}{\boldsymbol{\Delta}}
\newcommand{\boldeta}{\boldsymbol{\eta}}
\newcommand{\bxi}{\boldsymbol{\xi}}
\newcommand{\bGamma}{\boldsymbol{\Gamma}}
\newcommand{\bSigma}{\boldsymbol{\Sigma}}
\newcommand{\balpha}{\boldsymbol{\alpha}}
\newcommand{\bOmega}{\boldsymbol{\Omega}}
\newcommand{\btheta}{\boldsymbol{\theta}}
\newcommand{\bmu}{\boldsymbol{\mu}}
\newcommand{\bnu}{\boldsymbol{\nu}}
\newcommand{\bgamma}{\boldsymbol{\gamma}}

\newcommand{\bse}{\begin{eqnarray*}}
\newcommand{\ese}{\end{eqnarray*}}
\newcommand{\be}{\begin{eqnarray}}
\newcommand{\ee}{\end{eqnarray}}
\newcommand{\bsq}{\begin{equation*}}
\newcommand{\esq}{\end{equation*}}
\newcommand{\bq}{\begin{equation}}
\newcommand{\eq}{\end{equation}}
\newcommand{\var}{\mbox{var}}
\newcommand{\trace}{\hbox{trace}}
\newcommand{\wh}{\widehat}
\newcommand{\wt}{\widetilde}
\newcommand{\eff}{_{\rm eff}}
\newcommand{\sub}{{\rm sub}}
\newcommand{\cat}{{\rm cat}}
\newcommand{\eLL}{\mathcal L}
\newcommand{\n}{\nonumber}
\newcommand{\bias}{\mbox{bias}}
\newcommand{\vecl}{\mbox{vecl}}
\newcommand{\AIC}{\mbox{AIC}}
\newcommand{\BIC}{\mbox{BIC}}
\newcommand{\MSE}{\mbox{MSE}}
\newcommand{\rank}{\mbox{rank}}
\newcommand{\cov}{\mbox{cov}}
\newcommand{\corr}{\mbox{corr}}
\newcommand{\argmin}{\mbox{argmin}}
\newcommand{\argmax}{\mbox{argmax}}
\newcommand{\diag}{\mbox{diag}}
\newcommand{\trans}{^{\rm \top}}
\newcommand{\bTheta}{\boldsymbol\Theta}
\newcommand{\bta}{\boldsymbol\eta}
\newcommand{\bphi}{\boldsymbol\phi}
\newcommand{\btau}{\boldsymbol\tau}
\newcommand{\boeta}{\boldsymbol\eta}
\newcommand{\bpsi}{\boldsymbol\psi}
\newcommand{\0}{{\bf 0}}
\newcommand{\A}{{\bf A}}
\newcommand{\U}{{\bf U}}
\newcommand{\V}{{\bf V}}
\newcommand{\e}{{\bf e}}
\newcommand{\R}{{\bf R}}
\newcommand{\G}{{\bf G}}
\newcommand{\bO}{{\bf O}}
\newcommand{\B}{{\bf B}}
\newcommand{\D}{{\bf D}}
\newcommand{\K}{{\bf K}}
\newcommand{\g}{{\bf g}}
\newcommand{\f}{{\bf f}}
\newcommand{\h}{{\bf h}}
\newcommand{\I}{{\bf I}}
\newcommand{\M}{\mbox{ $\mathcal{M}$}}
\newcommand{\BB}{\mbox{ $\mathcal{B}$}}
\newcommand{\N}{\mbox{ $\mathcal{N}$}}
\newcommand{\T}{{\bf T}}
\newcommand{\bP}{{\bf P}}
\newcommand{\s}{{\bf s}}
\newcommand{\m}{{\bf m}}
\newcommand{\W}{{\bf W}}
\newcommand{\w}{{\bf w}}
\newcommand{\X}{{\bf X}}
\newcommand{\x}{{\bf x}}
\newcommand{\tx}{{\widetilde \x}}
\newcommand{\Y}{{\bf Y}}
\newcommand{\C}{{\bf C}}
\newcommand{\tY}{{\widetilde Y}}
\newcommand{\y}{{\bf y}}
\newcommand{\Z}{{\bf Z}}
\newcommand{\z}{{\bf z}}
\newcommand{\Ybar}{{\overline{Y}}}
\newcommand{\Xbar}{{\overline{\X}}}
\newcommand{\xbar}{{\overline{\x}}}
\newcommand{\wbar}{{\overline{\W}}}
\newcommand{\bSig}{{\bf \Sigma}}
\newcommand{\bLam}{{\bf \Lambda}}
\def\th{^{th}}
\def\S{{\bf S}}
\def\L{{\bf L}}
\def\u{{\bf u}}
\def\v{{\bf v}}
\def\T{{\bf T}}
\def\bO{{\bf O}}
\def\I{{\bf I}}
\def\K{{\bf K}}
\def\t{{\bf t}}
\def\b{{\bf b}}
\def\r{{\bf r}}
\def\V{{\bf V}}
\def\c{{\bf c}}
\def\a{{\bf a}}
\def\vec{\mbox{vec}}

\newcommand{\cvec}[1]{{\mathbf #1}}
\newcommand{\rvec}[1]{\vec{\mathbf #1}}
\newcommand{\minor}{{\rm minor}}
\newcommand{\spn}{{\rm Span}}
\newcommand{\range}{{\rm range}}
\newcommand{\mdiv}{{\rm div}}
\newcommand{\proj}{{\rm proj}}
\newcommand{\RR}{\mathbb{R}}
\newcommand{\NN}{\mathbb{N}}
\newcommand{\QQ}{\mathbb{Q}}
\newcommand{\ZZ}{\mathbb{Z}}
\newcommand{\EE}{\mathbb{E}}
\newcommand{\<}{\langle}
\renewcommand{\>}{\rangle}
\renewcommand{\emptyset}{\varnothing}
\newcommand{\attn}[1]{\textbf{#1}}
\newcommand{\bproof}{\bigskip {\bf Proof. }}
\newcommand{\eproof}{\hfill\qedsymbol}
\newcommand{\Disp}{\displaystyle}
\newcommand{\qe}{\hfill\(\bigtriangledown\)}
\newcommand*{\dif}{\mathop{}\!\mathrm{d}}

\newtheorem{thm}{Theorem}[section]
\newtheorem{lem}{Lemma}[section]
\newtheorem{rem}{Remark}[section]
\newtheorem{cor}{Corollary}[section]
\newcolumntype{L}[1]{>{\raggedright\let\newline\\\arraybackslash\hspace{0pt}}m{#1}}
\newcolumntype{C}[1]{>{\centering\let\newline\\\arraybackslash\hspace{0pt}}m{#1}}
\newcolumntype{R}[1]{>{\raggedleft\let\newline\\\arraybackslash\hspace{0pt}}m{#1}}

\newcommand{\tabincell}[2]{\begin{tabular}{@{}#1@{}}#2\end{tabular}}

\newtheorem{theorem}{Theorem}
\newtheorem{definition}{Definition}

\newcommand{\dT}{\top}

\newcommand{\algorithmicobs}{\textbf{Observations:}}
\newcommand{\algorithmicprior}{\textbf{Prior:}}
\newcommand{\PRIOR}{\item[\algorithmicprior]}
\newcommand{\OBS}{\item[\algorithmicobs]}

\newcommand{\algorithmicoutput}{\textbf{Output:}}
\newcommand{\OUTPUT}{\item[\algorithmicoutput]}

\newcommand{\whtheta}{\widehat{\theta}}
\newcommand{\htheta}{\hat{\theta}}
\newcommand{\UIP}{\mathrm{UIP}}
\newcommand{\pmean}{\mu}
\newcommand{\pvar}{\eta^2}
\newcommand{\UI}{\mathrm{I}_{\mathrm{U}}}
\newcommand{\MLE}{\mathrm{MLE}}
\newcommand{\init}{\mathrm{ini}}
\newcommand{\med}{\mathrm{med}}
\newcommand{\Var}{\mbox{Var}}

\setcitestyle{authoryear,open={(},close={)}}

\baselineskip=24pt
\begin{center}
{\Large \bf
Unit Information Prior for Adaptive Information
Borrowing from Multiple Historical Datasets
}
\end{center}

\vspace{2mm}
\begin{center}
{\bf Huaqing Jin and Guosheng Yin$^{*}$ }
\end{center}

\begin{center}

Department of Statistics and Actuarial Science\\
The University of Hong Kong\\
Pokfulam Road, Hong Kong\\

\vspace{2mm}

{\em *Correspondence}: gyin@hku.hk\\

\end{center}
\noindent{Abstract.}
In clinical trials, there often exist multiple historical studies for
the same or related treatment investigated in the current trial.
Incorporating historical data in the analysis of the current study is of great importance,
as it can help to gain more information, improve efficiency, and
provide a more comprehensive evaluation of treatment.
Enlightened by the unit information prior (UIP) concept in the reference Bayesian test,
we propose a new informative prior called UIP from an information perspective
that can adaptively
borrow information from multiple historical datasets. We consider
both binary and continuous data and
also extend the new UIP methods to linear regression settings.
Extensive simulation studies demonstrate that our method is comparable
to other commonly used informative priors,
while the interpretation of UIP is intuitive and
its implementation is relatively easy.
One distinctive feature of UIP is that its construction
only requires summary statistics commonly reported in the literature
rather than the patient-level data.
By applying our UIP methods to phase III clinical trials
for investigating the efficacy of memantine in Alzheimer's disease,
we illustrate its ability of adaptively borrowing information from multiple
historical datasets in the real application.

\vspace{0.5cm}

\vspace{1cm}
\section{Introduction}
	
Prior distributions are crucial in Bayesian data analysis and inference.
By incorporating many sources of knowledge, such as expert opinions and historical data,
a properly elicited prior distribution can help to analyze
the current data more efficiently and thus reduce the cost and ethical hazard in clinical trials.
With a large number of subjects followed for a long period of time,
large-scale clinical trials are typically expensive and suffer from high ethical risk.
Clinical trials are rarely conducted in isolation, and
thus it is critical to combine multiple samples systematically to improve
the analysis of the current study.
For example, several cancer clinical trials on the
same or similar types of treatment are often conducted
with patients of different ethnicity groups or
disease sub-types and sometimes in different countries \citep{brahmer2015nivolumab,borghaei2015nivolumab,wu2019nivolumab}.
These trials typically have comparable settings, e.g., using the same endpoints with
similar follow-ups and eligibility criteria,
hence information can be borrowed across them to improve efficiency.
It may also happen that trials investigate different treatments,
while the same control arm is used \citep{borghaei2015nivolumab,rittmeyer2017atezolizumab}, and
thus the data in the control arm are valuable to the current trial
as a common benchmark for comparison.

A major challenge associated with multiple
historical datasets is to determine the amount of information borrowed
from each dataset for the current study.
Differences in patient populations or other trial-specific settings lead
to heterogeneity among the current trial and historical trials,
and thus the full-borrowing strategy is typically imprudent,
as it would inflate the type I error substantially.
Moreover, when the total sample size of the historical datasets is large,
the information from historical datasets might overwhelm the analysis of the current study
which is not desirable,
as the data from the current study should typically dominate the analysis.

Several methods have been proposed for adaptively borrowing information from historical data.
\cite{pocock1976combination}
considers the difference in the parameters of interest between the current and historical datasets
and models this difference with a zero-mean random variable.
\cite{ibrahim2000power} propose the joint power prior (JPP) to
discount the historical dataset by a power parameter in the range of $[0,1]$.
\cite{duan2006evaluating} and \cite{neuenschwander2009note} modify the JPP by adding a normalization term,
which is referred to as the modified power prior (MPP).
\cite{banbeta2019modified} and \cite{gravestock2019power} further extend the power prior to
multiple historical datasets with binary endpoints.
Using a hierarchical model for the between-trial heterogeneity,
\cite{neuenschwander2010summarizing} develop the meta-analytic-predictive (MAP) prior via
deriving the predictive distribution of the model parameter resulting from the analysis of the historical datasets.
To further account for the prior-data conflict,
\cite{schmidli2014robust} make the MAP prior more robust
by incorporating a noninformative component in the prior distribution \citep{weber2019applying}
which is called robust MAP (rMAP).
\cite{hobbs2011hierarchical} propose the commensurate prior by
using the commensurate parameter directly to parameterize the commensurability
between the historical and current data.
There are also shrinkage estimation methods to borrow information from
historical data \citep{rover2020bounds,rover2020dynamically},
which are under the hierarchical model framework.
All the aforementioned methods can borrow information
according to the consistency between the historical and current datasets.
Nonetheless, Pocock's method and the commensurate prior are typically
applicable to the case with a single historical dataset.
When multiple historical datasets exist,
the naive extensions for both methods do not take the
underlying correlation among historical datasets into consideration.
For the commensurate prior,
when multiple historical datasets are involved,
the formula under the non-Gaussian case would be complicated which causes difficulty in practice and
the interpretations of multiple commensurate parameters may not be intuitive
as the commensurability concept defined by \cite{hobbs2011hierarchical} is typically
for the case with a single historical dataset.
The MAP prior relies on the exchangeability assumption and
adopts a single parameter to parameterize the heterogeneity between
the current and historical datasets.
With a single parameter, the relative contributions of historical datasets
are not accounted for, i.e.,
heterogeneity of the coherence between the current and each historical dataset
cannot be incorporated.

\cite{kass1995reference}
use the Fisher information to define the amount of information and
set the amount of information in the prior equal to that of a single
observation to conduct Bayesian hypothesis tests using Bayes factors.
Motivated by the information in a single observation,
we propose the unit information prior (UIP) as a new class of informative prior
distributions to dynamically borrow information from multiple historical datasets.
Unlike other methods which are constructed from the likelihood function of historical datasets,
the UIP is constructed from the information perspective and
parameterizes the amount of Fisher information in the prior distribution directly.
The amount of information in the UIP is closely related to the
effective sample size (ESS) defined by \cite{morita2008determining}, and
thus the UIP framework can straightforwardly control the ESS in the prior distribution.
Moreover, our method considers the heterogeneity between the current and historical datasets separately
which guarantees the efficiency of information borrowing.
As it is elicited based on the Fisher information,
often it only requires summary statistics of the historical data commonly reported
in publications (e.g., point estimates and 95\% confidence intervals) rather
than the patient-level data.
The UIP is easy to implement with multiple historical datasets,
whose parameters have intuitive interpretations.

The rest of this paper is organized as follows.
In Section~\ref{sec:methodology}, we introduce the general framework of the UIP methods.
In Section~\ref{sec:example}, we illustrate the UIP framework under the
single-arm trial with binary and continuous data respectively, and
discuss its connection with the power prior, commensurate prior and MAP prior
in terms of the conditional prior distribution
as well as making an extension of the UIP to linear models.
We also discuss the effective sample size (ESS) in the UIP \citep{morita2008determining,morita2012prior}.
Extensive simulation studies are presented in Section~\ref{sec:simulation} where
we demonstrate the dynamic borrowing property of the UIP,
and compare different priors
for multiple historical datasets.
Section~\ref{sec:realdata} provides an example from six phase III clinical trials for Alzheimer's disease which
illustrates the behavior of our UIP approach in the real data application.
We conclude the article with a brief discussion in Section~\ref{sec:discussion}.
	
	\section{Unit Information Prior}\label{sec:methodology}
Let $D=\{Y_{1}, \ldots, Y_{n}\}$
denote the data of the current trial of sample size $n$.
Suppose that there are $K$ historical datasets $\{D_1, \ldots, D_K\}$ related to the current study,
where $D_k=\{Y_{k,1}, \ldots, Y_{k, n_k}\}$ denotes the $k$th dataset of
size $n_k$, for $k =1, \ldots, K$.
The parameter of interest is often the treatment effect, denoted by $\theta$,
for the current study,
while the counterpart of $\theta$ for $D_k$ is denoted as $\theta_k$ for $k=1, \ldots, K$.
The likelihood function of $\theta_k$ based on $D_k$ is denoted
by $L^{(k)}(\theta_k|D_k)$.

The UIP is constructed directly from an information perspective
under the normal approximation.
When eliciting an informative prior for the parameter of interest $\theta$,
we are mainly interested in the accuracy and precision,
i.e., the correctness
and the amount of information contained in the prior distribution.
Under the normal approximation, the accuracy of the prior distribution is determined
by the mean of the prior
and the amount of the Fisher information in the prior is the inverse of the variance.
Thus, the UIP framework focuses on
the mean and variance of the prior distribution,
and the amount of information in the prior distribution can be
explicitly controlled.
Moreover, as UIP only requires the first and second moments,
often the summary statistics of the historical data (e.g., mean and variance)
would be sufficient to derive the
prior distribution, which is the typical case
in practice as the patient-level historical datasets
are not easy to obtain.

When considering information borrowing from historical datasets,
the parameter of interest $\theta$ is assumed to be close to
the counterpart of historical datasets $\theta_k$.
Due to heterogeneity among historical datasets,
we introduce a weight parameter $w_k$ for the historical dataset $D_k$
to measure the relative closeness between the current dataset $D$ and the historical one $D_k$.
The mean of the prior is defined as
 $\sum_{k=1}^K w_k \theta_k$,
with the weight parameter $w_k \in (0, 1)$ and $\sum_{k=1}^K w_k = 1$,
where $w_k$
can be also viewed as the measurement of the relative contribution from dataset $D_k$
to the analysis of the current study.
The larger value of $w_k$, the more contribution from $D_k$.

Following the work of \cite{kass1995reference}, we adopt the Fisher information
as the measurement of the amount of information in the data.
As a result, we define the unit information (UI) for $\theta_k$ in the dataset $D_k$ as
\bse
\UI(\theta_k) = -\frac{1}{n_k}\frac{\partial^2 \log L^{(k)}(\theta_k|D_k)}{\partial \theta_k^2},
\ese
i.e., $\UI(\theta_k)$ is the observed Fisher information of $D_k$ averaged at a unit observation level.
When taking the heterogeneity of the historical datasets into consideration,
the contribution of each historical dataset to the current study should be distinct.
Therefore, the unit information from all the $K$ historical datasets
is defined as $\sum_{k=1}^K w_k\UI(\theta_k)$.
We introduce $M$ as the sample size of the total amount of information borrowed from the
$K$ datasets,
and then the amount of information contained in the prior is
$M\sum_{k=1}^K w_k\UI(\theta_k)$.
Under the normal approximation, the variance of the prior distribution is
$\left\{M\sum_{k=1}^K w_k\UI(\theta_k)\right\}^{-1}$.

Therefore, to adaptively borrow information from different datasets,
we formulate the UIP framework as follows,
\bse
\theta|(M, w_1, \ldots, w_K, D_1, \ldots, D_K)
&\sim& \pi(\theta|M, w_1, \ldots, w_K, D_1, \ldots, D_K),  \\
\mathrm{with}  \quad
\EE_{\pi}(\theta) &=& \sum_{k=1}^K w_k \theta_{k}, \\
\Var_{\pi}(\theta) &=& \left\{M\sum_{k=1}^K w_k \UI(\theta_{k})\right\}^{-1}.
\ese
As $\theta_k$ is typically unknown, we adopt the maximum likelihood estimator
(MLE) $\hat{\theta}_{k}$ to replace $\theta_k$
while keeping $M, w_1, \ldots, w_K$
as unknown parameters that are determined adaptively by the data.
We choose the MLE as the estimate of $\theta_k$
for its desirable properties \citep{hunter2014notes}, e.g., consistency and asymptotic efficiency.
Moreover, in many cases, the MLE is simply the mean of the samples which can be easily obtained.

The specific form of $\pi$ depends on the parameter of interest $\theta$.
For example, if $\theta$ is the mean parameter of a normal distribution,
the UIP of $\theta$ can have a normal distribution;
and if $\theta$ is the rate parameter of a Bernoulli distribution,
the UIP of $\theta$ may conform to a Beta distribution.

The parameter $M$ determines the total number of units corresponding to
the amount of information borrowed from all historical datasets as a whole.
It is shown that the value of $M$ is closely related to the ESS defined by \cite{morita2008determining}.
In practice, $M$ can be either fixed
when we aim to control the amount of information borrowed from historical data
or unknown by setting a hyper-prior on $M$.
When there is a lack of prior information on $M$,
a noninformative uniform prior is recommended as
it is a standard practice in Bayesian methods;
otherwise, Gamma or Poisson distribution can be used
to incorporate the prior information of $M$ in the hyper-prior.
For the simulation studies, unless otherwise specified,
we set a Uniform$(0, \sum_{k=1}^K n_k)$ as the hyper-prior for $M$, i.e.,
the value of $M$ cannot exceeds the combined sample size of all historical datasets.
If the total sample size of historical datasets is too large
compared to the sample size of the current study,
it is suggested to set the upper bound of Uniform distribution at $n$,
the sample size of the current dataset.

It is essential to determine the values of weight parameters which
characterize the amount of information borrowed from each historical dataset.
The values of $w_k$'s reveal the relative importance of historical datasets.
If one historical dataset is more consistent with the current study compared with
the others, the corresponding weight parameter should be larger.
Furthermore, the absolute weight $Mw_k$ can be interpreted as the number of units
of information
borrowed from historical dataset $D_k$.
Therefore, the amount of information borrowing from different
datasets mainly relies upon the weight parameters,
and the total amount of information borrowed from all historical datasets
is determined by $M$.

We propose two approaches to determining the values of the weight parameters.
One is a fully Bayesian approach by imposing a hyper-prior on $(w_1, \ldots, w_K)$.
As all the values of $w_k$'s are between $0$ and $1$ satisfying
the constraint $\sum_{k=1}^Kw_k=1$,
it is natural to assign a Dirichlet prior to $(w_1, \ldots, w_K)$.
We take the sample sizes of historical datasets into consideration
by selecting suitable parameters for the Dirichlet distribution.
Intuitively,
it is preferable to assign a higher weight to the historical dataset with larger sample size,
while we should also prevent
a historical dataset with extremely large sample size from
dominating the information borrowing.
To strike a balance,
we recommend to set the hyper-prior for $(w_1, \ldots, w_K)$
as Dirichlet$(\gamma_1, \ldots, \gamma_K)$ where $\gamma_k = \min(1, {n_k}/{n})$.
We refer to the UIP with a Dirichlet prior distribution as UIP-Dirichlet.

The other approach is to first measure the distances between the current
dataset and the historical ones.
To determine the values of weight parameters,
a proper measure of the ``distance'' between two datasets is needed
for measuring their similarity.
The Jensen--Shannon (JS) divergence is a commonly used metric for measuring
the dissimilarity between two probability distributions
\citep{itzkovitz2010overlapping,sims2009alignment,goodfellow2014generative}.
An alternative
is the Kullback--Leibler (KL) divergence,
while we adopt the JS  divergence
due to its symmetry property.

Similar to the discussion in the UIP-Dirichlet method,
the weight parameter $w_k$ should tend to be small when the sample size of $D_k$ is small,
while $w_k$ should not be too large even if the sample size of $D_k$
is extremely large.
The JS divergence automatically penalizes the relatively small sample size
(compared with the current dataset) of the historical dataset.
When the sample size of $D_k$ is larger than that of the current dataset,
we randomly select $n$ samples from $D_k$ to calculate the JS divergence
and repeat this procedure for a large number of times to obtain the average.
More specifically, the weight parameters are determined as follows.
\begin{enumerate}[(1)]
    \item Specify an initial non-informative prior
    (e.g., Jeffreys prior) for the parameter
    $\theta$ under the current dataset $D$ and $\theta_k$ under each $D_k$.
    Based on the initial prior, we obtain the initial posteriors  $f_{\init}(\theta|D)$.

   \item For $k=1, \ldots, K$, when $n_k \le n$, obtain
   the initial posterior $f_{\init}(\theta|D_k)$ and calculate the JS divergence
   \bse
   d_k=\mathrm{JS}(D|D_k)=\frac{\mathrm{KL}(D|D_k)+ \mathrm{KL}(D_k|D)}2,
   \ese
   where $\mathrm{KL}(D|D_k)$ represents the KL divergence between two density functions $f_{\init}(\theta|D)$ and $f_{\init}(\theta|D_k)$,
$$  \mathrm{KL}(D|D_k)=\int f_{\init}(\theta|D)\log
\left\{\frac{f_{\init}(\theta|D)}{f_{\init}(\theta|D_k)}\right\}\dif \theta.
$$
    When $n_k > n$, randomly draw $n$ samples from $D_k$ for $N$ times to obtain $\{D_k^{(l)}\}_{l=1}^N$ and compute the initial posteriors $\{f_{\init}(\theta|D_k^{(l)})\}_{l=1}^N$, then
    calculate the distance $d_k$ as the average ${\sum_{l=1}^N \mathrm{JS}(D|D_k^{(l)})}/{N}$.
    \item The weight parameters are defined as
    $$w_k= \frac{1/d_k}{\sum_{s=1}^K(1/d_s)}$$ for $k=1, \ldots, K$.
\end{enumerate}
In extremely rare cases, $D$ and $D_k$ may be exactly the same which results in the zero JS divergence.
As a remedy, we add a small number, say $10^{-6}$,  to the $d_k$ to avoid the division-by-zero problem.
We refer to the UIP in conjunction with the JS divergence as UIP-JS, where
the weights are prespecified and the only unknown parameter is $M$ in the UIP-JS.

It is also possible to use other methods (e.g., the empirical Bayes method)
to pre-determine the weight parameters before sampling.
However, in terms of the computation,
the JS divergence is easier compared with the empirical Bayes method.
Moreover, the JS divergence measures the dissimilarity between two datasets
from an information perspective,
which is consistent with the UIP framework for the prior elicitation.


\section{UIP with Binary and Continuous Data} \label{sec:example}
We illustrate the UIP methods in a single-arm trial with continuous and binary data
respectively.
We also discuss the connections among the power prior, commensurate prior, MAP prior and UIP
for continuous data in terms of the conditional prior distribution,
as well as extending our UIP to linear models.
Moreover, the relationship between the amount parameter $M$ and
the ESS of the informative prior distribution is investigated.

\subsection{Continuous Data}
Suppose that $\{Y_1, \ldots, Y_n\}$ are independent
and identically distributed (i.i.d.) from $N(\theta, \sigma^2)$
and $\{Y_{k, 1}, \ldots, Y_{k, n_k}\}$ are i.i.d.
from $N(\theta_k, \sigma^2_k)$ for $k=1, \ldots, K$.
The parameter of interest is the mean $\theta$ and
the unit information for $D_k$ evaluated at the corresponding MLE
$\htheta_{k}$ (which is the sample mean) is
\bse
\UI(\htheta_{k}) = \frac{1}{\sigma^2_{k}}.
\ese
We impose an inverse-Gamma prior for $\sigma^2$, e.g.,
$\sigma^2 \sim \mathrm{InvGa}(0.01, 0.01)$.
If the calculation involves $\sigma^2_k$, we simply replace $\sigma^2_k$ with
its MLE $\hat{\sigma}^2_{k}$. As a result,
we obtain the UIP as
\bse
\theta|(M, w_1, \ldots, w_K, D_1, \ldots, D_K) &\sim& N(\pmean, \pvar),
\ese
where
\bse
\pmean &=& \sum_{k=1}^K w_k \htheta_{k}, \\
\pvar &=& \left(M\sum_{k=1}^K \frac{w_k}{\hat{\sigma}^2_{k}}\right)^{-1}.
\ese

Under the normal distribution setting, the MPP, local commensurate prior (LCP),
MAP prior and UIP are closely related as shown in formula~(\ref{eq:4priors}).
We extend the MPP to the case with multiple historical datasets as follows,
\bse
\pi^{\mathrm{MPP}}(\theta|\alpha_1, \ldots, \alpha_K, D_1, \ldots, D_K)
&\propto& \pi^{\init}(\theta)\prod_{k=1}^K L^{(k)}(\theta|D_k)^{\alpha_k},\\
\alpha_k &\sim& \mathrm{Beta}(a, b), \  k = 1, \ldots, K, \\
\sigma^2 &\sim& \mathrm{InvGa}(\zeta, \zeta), \\
\pi^{\init}(\theta) &\propto& 1,
\ese
where $\alpha_k$ is the power parameter for $D_k$ and $\pi^{\init}$ is the initial prior.

We also extend the LCP method to the case with $K$ historical datasets.
Denoting the commensurate parameter for dataset $D_k$ by $\tau_k$, the LCP is given by
\bse
\pi^{\mathrm{LCP}}(\theta|\tau_1, \ldots, \tau_K, D_1, \ldots, D_K) &\propto&
\pi^{\init}(\theta)\prod_{k=1}^K \int L^{(k)}(\theta_k|D_k)
N\big(\theta; \theta_k, \frac{1}{\tau_k}\big)\dif\theta_k, \\
\log(\tau_k) &\sim& \mathrm{Unif}(\xi_1, \xi_2), \ k = 1, \ldots, K,\\
\sigma^2 &\sim& \mathrm{InvGa}(\zeta, \zeta), \\
\pi^{\init}(\theta) &\propto& 1.
\ese

Denoting the between-trial standard deviation as $\tau$,
the MAP prior with multiple historical datasets can be written as
\bse
\theta_1, \ldots, \theta_K, \theta |\mu^{\mathrm{MAP}},\tau
&\sim& N(\mu^{\mathrm{MAP}}, \tau^2), \\
\pi(\mu^{\mathrm{MAP}}) &\propto&  1, \\
\tau &\sim& \mathrm{HN}(0, 1),
\ese
where $\mathrm{HN}$ denotes the half-normal distribution.

The MPP and LCP respectively use the power parameter $\alpha_k$ and the commensurate parameter $\tau_k$
as the measurement of consistency between the current dataset and
each historical dataset $D_k$.
The MAP prior adopts the exchangeability assumption and utilizes
a single between-trial dispersion parameter $\tau$ to measure the
heterogeneity among the current and historical datasets.
In the UIP methods, the weight parameters measure the relative consistency
of the historical datasets with respect to the current dataset,
while the total amount parameter $M$ measures the overall information
borrowed from historical data.

Given the corresponding hyper-parameters,
the specific forms of the MPP, LCP, MAP and UIP methods for continuous data are
given as follows:
\be\label{eq:4priors}
\pi^{\mathrm{MPP}}(\theta|\alpha_1, \ldots, \alpha_K, D_1, \ldots, D_K)
&\sim&
N\left(\sum_{k=1}^K \frac{\alpha_kn_k/\hat{\sigma}_k^2}{\sum_{s=1}^K \alpha_sn_s/\hat{\sigma}^2_s} \htheta_{k},
\left(\sum_{k=1}^K \alpha_k n_k \frac{1}{\hat{\sigma}^2_k}\right)^{-1}
\right), \nonumber \\ \nonumber
\pi^{\mathrm{LCP}}(\theta|\tau_1, \ldots, \tau_K, D_1, \ldots, D_K)
&\sim&
N\left(
\sum_{k=1}^K \frac{n_k\tau_k/(\tau_k\hat{\sigma}^2_k+n_k)}{\sum_{s=1}^K n_s\tau_s/(\tau_s\hat{\sigma}^2_s+n_s)} \htheta_{k},
\left(
\sum_{k=1}^K \frac{n_k\tau_k\hat{\sigma}_k^2}{\tau_k\hat{\sigma}^2_k+n_k}\frac{1}{\hat{\sigma}_k^2}
\right)^{-1}
\right), \\ \nonumber
\pi^{\mathrm{MAP}}(\theta|D_1, \ldots, D_K, \tau)
&\sim& N\left(
\sum_{k=1}^K
\frac{n_k/(\hat{\sigma}_k^2 +n_k\tau)}
{\sum_{s=1}^Kn_s/(\hat{\sigma}_s^2 +n_s\tau)}\htheta_k,
\frac{1}{\sum_{k=1}^Kn_k/(\hat{\sigma}_k^2 +n_k\tau)}+\tau^2\right) \\
\pi^{\mathrm{UIP}}(\theta|M, w_1, \ldots, w_K, D_1, \ldots, D_K)
&\sim& N\left(
 \sum_{k=1}^K w_k \htheta_{k},
\left(\sum_{k=1}^K w_k M \frac{1}{\hat{\sigma}_k^2}\right)^{-1}
\right).
\ee
Conditioning on the hyper-parameters,
the MPP, LCP, MAP and UIP approaches all lead to normal distributions
with different means and variances. Interestingly,
the means of all four priors can be written in a form of a weighted sum of
the individual sample means.
The UIP adopts weight parameters in a much more direct way,
while the MPP parameterizes weights as an increasing function of the power parameters.
The LCP method utilizes commensurate parameters $\tau_k$ to measure
the commensurability between the current and historical datasets, where a
larger value of $\tau_k$ indicates a higher level of commensurability,
and the weight naturally increases with $\tau_k$.
The MAP prior utilizes a single dispersion parameter $\tau$ to control the information borrowing,
thus weight parameters mainly rely on the historical variances
other than the dispersion parameter $\tau$.

The precision (i.e., the inverse of the variance) of the four priors can be used to quantify the
amount of information borrowed from historical datasets.
In particular, the precision of the MPP, LCP and UIP methods can be written as a weighted sum of the
observed Fisher information from each dataset.
For the MPP method, the number of units of information for $D_k$ is determined
by the product of the corresponding power parameter $\alpha_k$ and the sample size $n_k$.
The LCP approach adopts an increasing function of $\tau_k$ and $n_k$
as the amount of information borrowed from $D_k$.
The amount of information from $D_k$ under the UIP framework corresponds to the
product of the weight parameter $w_k$ and the total amount parameter $M$,
which has the most transparent form and intuitive interpretation.
Under the exchangeability assumption,
the MAP prior borrows information from multiple historical datasets as a whole where
the amount of information borrowed is a decreasing function of the dispersion parameter $\tau$.

\subsection{Binary Data}
Suppose that $\{Y_1, \ldots, Y_n\}$ are i.i.d. samples from $\mathrm{Bernoulli(\theta)}$,
while $\{Y_{k, 1}, \ldots, Y_{k, n_k}\}$ are those from $\mathrm{Bernoulli}(\theta_k)$ for $k=1, \ldots, K$.
The UI under the $k$th historical dataset $D_k$ evaluated at $\hat{\theta}_{k}$ is
\bse
\UI(\hat{\theta}_{k})=\frac{1}{\htheta_{k}(1-\htheta_{k})}.
\ese
As the support of the rate parameter $\theta$ is $[0, 1]$,
we assign a Beta prior on $\theta$.
Denote the prior mean and prior variance of UIP as $\pmean=\sum_{k=1}^K w_k \htheta_{k}$
and $\pvar=\left\{M\sum_{k=1}^K {w_k}/\{\htheta_k(1-\htheta_k)\}\right\}^{-1}$.
Thus, the UIP of $\theta$ can be written as
\bse
\theta|(M, w_1, \ldots, w_K, D_1, \ldots, D_K) &\sim& \mathrm{Beta}(\alpha, \beta),
\ese
where the two Beta distribution parameters can be easily derived by solving the mean and
variance equations,
\bse
\alpha &=& \pmean \left\{\frac{\pmean(1-\pmean)}{\pvar} - 1 \right\},\\
\beta &=& (1-\pmean) \left\{\frac{\pmean(1-\pmean)}{\pvar} - 1 \right\}.
\ese
%

\subsection{Linear Regression}
Under a linear regression model,
$Y_i \sim N(\x_i\trans\bbeta, \sigma^2)$, where $\bbeta=(\beta_0, \ldots, \beta_p)\trans$ is the regression coefficients and $\x_i=(1, x_{1, i}, \ldots, x_{p, i})\trans$ is the covariate vector associated with the outcome $Y_i$.
For the $k$th historical dataset,
$Y^{(k)}_{i} \sim N(\x_i^{(k)\top}\bbeta^{(k)}, \sigma_k^2)$,
where $\bbeta^{(k)}=(\beta_0^{(k)}, \ldots, \beta_{p}^{(k)})\trans$ and
$\x^{(k)}_i=(1, x_{1, i}^{(k)}, \ldots, x_{p, i}^{(k)})\trans$ for $k=1,\ldots,K$ and $i=1, \ldots, n_k$.

Under the linear model, we obtain the UI for $D_k$ evaluated at
$\hat{\beta}^{(k)}_{l}$ as
\be \label{lmui}
\UI(\hat{\beta}^{(k)}_{l}) =
\frac{1}{n_k {\rm Var}(\hat{\beta}_l^{(k)})},
\ee
where $\hat{\beta}^{(k)}_{l}$ is the MLE of $\beta_l^{(k)}$ and
${\rm Var}(\hat{\beta}_l^{(k)})$ is the corresponding variance
for $l=0, \ldots, p$.
Thus, the UIPs of the regression coefficients are given by
\bse
\beta_l|(M, w_1, \ldots,  w_K, D_1, \ldots, D_K) &\sim&
N\left(\sum_{k=1}^K w_k\hat{\beta}^{(k)}_{l},
\Big\{M\sum_{k=1}^K w_k \UI(\hat{\beta}^{(k)}_{l}) \Big\}^{-1}\right),
\ese
for $l=0, \ldots, p$.

Certainly, we can assign the weight parameters and total amount parameter for each coefficient
individually, while this strategy involves too many unknown parameters, leading to
difficulties in the implementation of Markov chain Monte Carlo (MCMC).
Hence, our parsimonious strategy is more desirable, i.e., we use the
same weights and $M$ for all coefficients. If not all regression
coefficients in the linear
model are shared between the current data and historical data,
we can impose UIPs on those shared coefficients only and
leave the unshared ones with non-informative priors.

\subsection{Prior Effective Sample Size} \label{subsec:ess}
For any prior distribution,
it is critical to quantify how much information is contained in the distribution
in terms of the effective sample size (ESS)
\citep{neuenschwander2010summarizing}.
In the sequel,
we discuss the ESS and its connection with the amount parameter $M$ in the UIP.

Given the weight parameters $w_k$ and the amount parameter $M$,
the ESS of the UIP can be easily obtained via the method of \cite{morita2008determining}.
For continuous data, the ESS is
$
\sigma^2M\sum_{k=1}^K (w_k/\sigma_k^2)
$.
In practice, when incorporating the historical information,
it is reasonable to assume $\sigma^2 \approx \sigma^2_1 \approx \cdots \approx \sigma^2_K$.
In such case, the ESS is approximately equal to the amount parameter $M$.
For binary data, the ESS is
\bse
\alpha + \beta = \frac{\mu(1-\mu)}{\eta^2} - 1 =
M \left\{\sum_{s=1}^K w_s \htheta_s \right\}
\left\{\sum_{l=1}^K w_l (1-\htheta_l) \right\}
\left\{\sum_{k=1}^K \frac{w_k}{\htheta_k(1-\htheta_k)}\right\}
- 1,
\ese
which is approximately equal to $M-1$  when $\htheta_1\approx \cdots \approx \htheta_K$.
Therefore, $M$ in the UIP methods
represents the amount of information contained in the informative prior.

It is also possible to obtain the ESS under the full Bayesian manner following
\cite{morita2012prior}.
To calculate the ESS under the full Bayesian manner,
we first define the $\epsilon$-information conditional prior $\pi_{\epsilon}(\theta|w_1,\ldots, w_K, M)$ such that it has the same mean but very large variance compared with
the conditional informative prior.
By integrating out the parameters $(w_1, \ldots, w_K, M)$, the marginal informative prior is
\bse
\pi(\theta|D_1, \ldots, D_K)=\int \pi(\theta|w_1,\ldots, w_K, D_1, \ldots, D_K, M) \pi(w_1, \ldots, w_K, M) \dif w_1 \cdots \dif w_K \dif M.
\ese
Suppose the expected dataset $\bar{D}^{(m)}$ contains $m$ samples
and all samples are $\bar{\theta}$ where $\bar{\theta}$ is the mean
of the distribution $\pi(\theta|D_1, \ldots, D_K)$.
This leads to the expected posterior as
\bse
\pi_{\epsilon}(\theta|\bar{D}^{(m)}) \propto L(\theta|\bar{D}^{(m)})
\int \pi_{\epsilon}(\theta|w_1,\ldots, w_K, M) \pi(w_1, \ldots, w_K, M) \dif w_1 \cdots \dif w_K \dif M.
\ese
The ESS is defined as the value of $m$ by minimizing
$
|\sigma_{\pi}^{-2} - \sigma^{-2}_{\epsilon}(\bar{D}^{(m)})|,
$
where $\sigma_{\pi}^2$ and $\sigma_{\epsilon}^2(\bar{D}^{(m)})$
are the variances under $\pi(\theta|D_1, \ldots, D_K)$ and $\pi_{\epsilon}(\theta|\bar{D}^{(m)})$, respectively.

\section{Simulation Studies} \label{sec:simulation}

We conduct extensive simulations to assess the characteristics of the UIP methods
with continuous data, and
the results for binary data are presented in the supplementary material.
First, we evaluate the ESS and the adaptive borrowing property of the UIP.
We then compare the UIP with Jeffreys prior, full-borrowing strategy,
MPP, LCP and rMAP priors under the single-arm trial scenario
in terms of the mean square error (MSE) as well as
hypothesis testing $H_0: \theta=\theta_0$ vs $H_1: \theta\neq \theta_0$.
The full-borrowing strategy refers to the analysis by directly pooling
the current and historical datasets and applying Jeffreys prior
for the pooled dataset.
For the MPP and LCP methods, we adopt flat initial priors,
i.e., $\pi^{\init}(\theta) \propto 1$.
We use a non-informative prior, $\mathrm{Beta}(1, 1)$, on
the power parameter $\alpha_k$ of the MPP method,
while a vague prior, $\mathrm{Uniform}(-30,30)$,
is imposed on the logarithm of the commensurate parameter $\log(\tau_k)$ for the LCP method.
Following \cite{neuenschwander2010summarizing}, the rMAP prior adopts the half-normal distribution
with scale parameter 1 for the dispersion parameter $\tau$ and assigns weight
$0.1$ to the noninformative component.

\subsection{Effective Sample Size} \label{simu:ess}
We justify the relationship between the amount parameter $M$ and ESS in the prior distribution.
For continuous data,
we adopt three historical datasets of sample sizes $(80, 100, 120)$ with
the sample size of the current dataset $100$.
The amount parameter $M$ varies from $50$ to $150$.
The mean $\theta$ and standard deviation $\sigma$ of the current dataset are fixed at $0$ and $1$,
while the means $(\theta_1, \theta_2, \theta_2)$ and standard deviations $(\sigma_1, \sigma_2, \sigma_3)$ of the historical datasets are randomly generated
from ranges $[-0.5, 0.5]$ and $[0.9, 1.1]$, respectively.
To obtain robust results, we repeat the experiment for $100$ times.

\begin{figure}[t]
\centering
\includegraphics[scale=0.6]{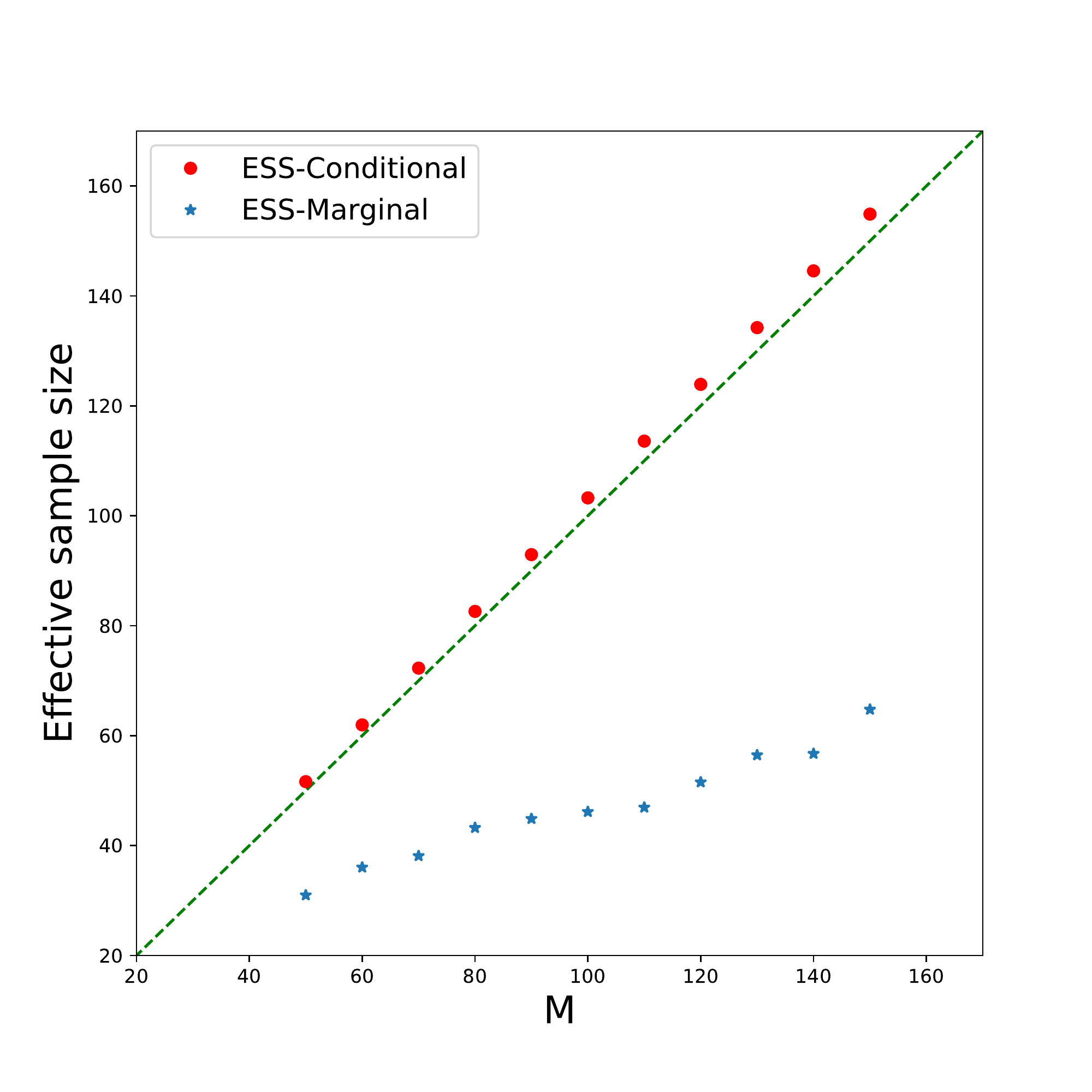}
\caption{
The effective sample size (ESS) of the
the conditional UIP and that of the marginal UIP for continuous data
when varying $M$ from $50$ to $150$. The dashed green line is the diagonal line.
}
\label{fig:ess}
\end{figure}

The ESS of the conditional prior distribution \citep{morita2008determining}
and that of the marginal prior distribution \citep{morita2012prior} under $100$ repetitions
are presented in Figure~\ref{fig:ess}, where the weights for the conditional UIP
are calculated via the JS method. For continuous data,
the ESS contained in the conditional UIP is close to the amount parameter $M$.
As the hyper-prior on weight parameters introduces more uncertainty,
the ESS in the marginal UIP decreases compared with that in the conditional UIP
given the same $M$ as desired.
However, the ESS of the marginal UIP still shows the same tendency with the amount parameter $M$,
i.e., it increases as $M$ increases.
The results for binary data show similar trends as shown in Web Figure~1 of the supplementary material.

\subsection{Adaptive Borrowing Property}\label{simu:abp}

We demonstrate that under the UIP, the values of the total amount parameter $M$,
the weight parameters $w_k$'s and the absolute weights $Mw_k$'s
can adapt to the level of consistency between
the historical datasets and current dataset.
If the historical datasets are close to the current, the UIP
borrows more information from historical datasets, i.e., the value of $M$ would be large.
When a certain dataset $D_k$ is more consistent with $D$ relative to other datasets,
more weight would be assigned to $D_k$.

\begin{figure}[t]
\centering
\begin{minipage}[t]{1\linewidth}
\centering
\includegraphics[scale=0.4]{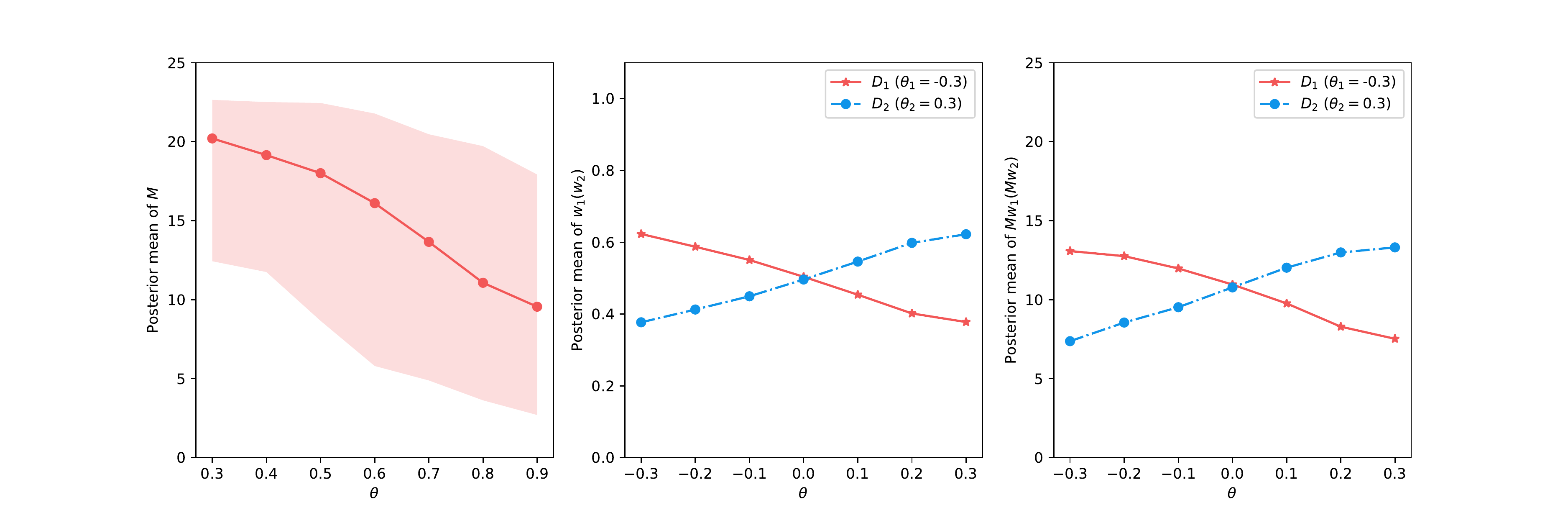}
\end{minipage}%

\begin{minipage}[t]{1\linewidth}
\centering
\includegraphics[scale=0.4]{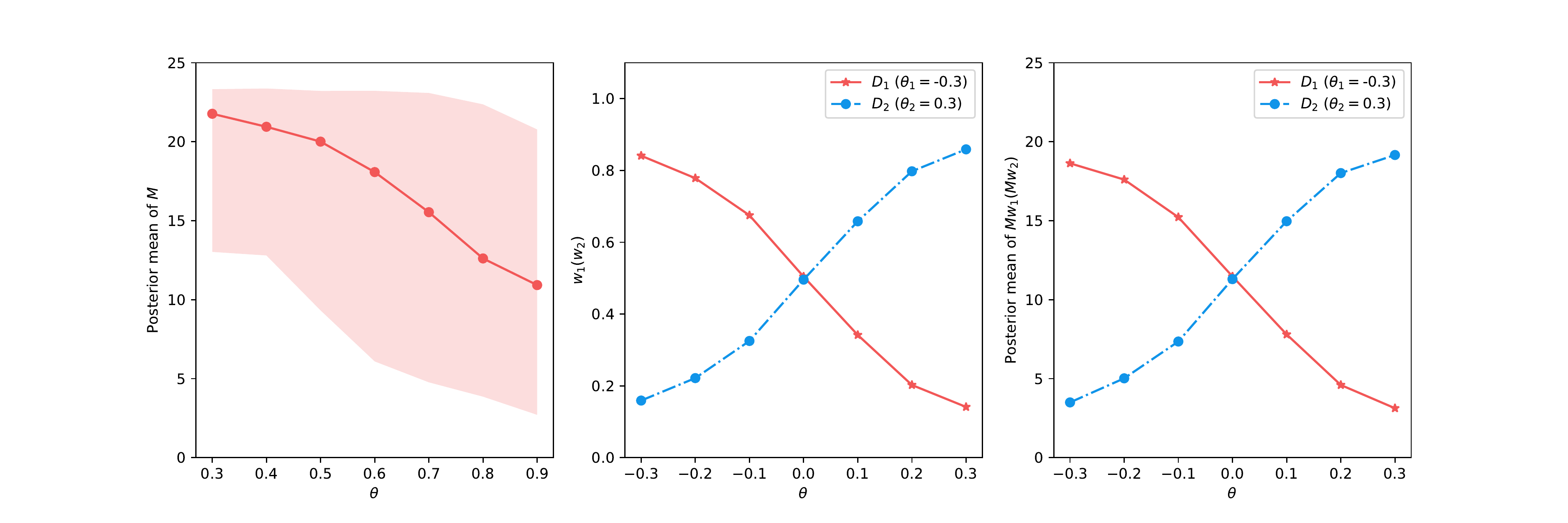}
\end{minipage}%
\caption{
Under the UIP-Dirichlet (top) and UIP-JS (bottom) methods
for continuous data with 100 repetitions,
when the population mean of the current dataset $D$ varies from $0.3$ to $0.9$,
the trend of $M$ (left panels) with
the red shadow area indicating the $95\%$ confidence interval and
when the population mean of the current dataset $D$ varies from $-0.3$ to $0.3$,
the trends of $w_1$ and $w_2$ (middle panels)
and the trends of $Mw_1$ and $Mw_2$ (right panels).
}
\label{fig:trendNorm}
\end{figure}

To examine the trend of the total amount parameter $M$,
we consider two historical datasets
$D_1$ and $D_2$ generated from $N(-0.3, 1)$ and $N(0.3, 1)$ respectively
with the same sample size $n_1=n_2=40$.
We vary $\theta$, the mean of the current dataset $D$ also with sample size $n=40$,
from $0.3$ to $0.9$ and fix the standard deviation at $\sigma=1$.
The hyper-prior for $M$ is set as $M \sim \mathrm{Uniform}(0, 40)$.
We draw the posterior samples of the total amount parameter $M$
under both UIP-Dirichlet and UIP-JS and
take the posterior mean of $M$ as the estimate.
We replicate the experiment for $100$ times, and
the averages of posterior means of $M$
under both priors are shown in the left column of Figure~\ref{fig:trendNorm}.
When the level of consistency between the
population mean of $D$ and those of historical datasets decreases,
the value of $M$ decreases,
indicating that less information is borrowed from historical datasets.

We further utilize two historical datasets
to investigate the trends of weight parameter $w_k$ and absolute weight $Mw_k$
for the UIP-Dirichlet and UIP-JS methods.
The historical data $D_1$ and
$D_2$ are drawn from $N(-0.3, 1)$ and $N(0.3, 1)$ respectively with sample sizes $n_1=n_2=40$
while the population mean of $D$ with $n=40$ varies from $-0.3$ to $0.3$ with
a fixed standard deviation $\sigma=1$.
The hyper-prior for $M$ is set as $M \sim \mathrm{Uniform}(0, 40)$.
We replicate the experiment for $100$ times to draw the plots
of the estimates of weight parameters $w_1$ ($w_2$) and absolute weights $Mw_1$ ($Mw_2$).
The right column of Figure~\ref{fig:trendNorm} demonstrates that
when the population mean of $D$ is closer to that of $D_1$,
$Mw_1$ is larger, indicating more information is borrowed from
$D_1$ compared with $D_2$;
and a similar trend is observed for $Mw_2$.
The tendency of the weight parameters $w_1 (w_2)$ is similar to
that of  $Mw_1 (Mw_2)$,
which indicates the change of $M$ is minor under this setting when the mean parameter $\theta$ varies.
It is reasonable because the overall level of consistency between the current and historical datasets
remains approximately the same
when varying $\theta$ between $\theta_1$ and $\theta_2$.
It is also worth noting that the heterogeneity of weight parameters under
the UIP-JS method is larger than that under the UIP-Dirichlet method.
For example, when the population mean of the current dataset $D$ is $0.3$,
the UIP-JS method assigns a weight of almost $0.8$ to the historical dataset $D_1$
(whose population mean is also $0.3$),
while the weight parameter of $D_1$ under
the UIP-Dirichlet method is around $0.6$, which is less extreme than that of UIP-JS.

The results for binary data are presented in Web Figure~2 of
the supplementary material, where similar phenomena can be observed.

\subsection{Single-Arm Trial Scenario}

We further compare our UIPs with Jeffreys prior, the full-borrowing method, MPP,
LCP and rMAP priors for continuous data.
We consider two historical datasets with sample sizes $n_1=100$ and $n_2=50$:
$D_1$ from $N(0.5, 1)$ and $D_2$ from $N(1, 1)$.
The variance for the current dataset $D$ is also fixed as $1$.

\begin{figure}[t]
\centering
\begin{minipage}[t]{1\linewidth}
\centering
\includegraphics[scale=0.4]{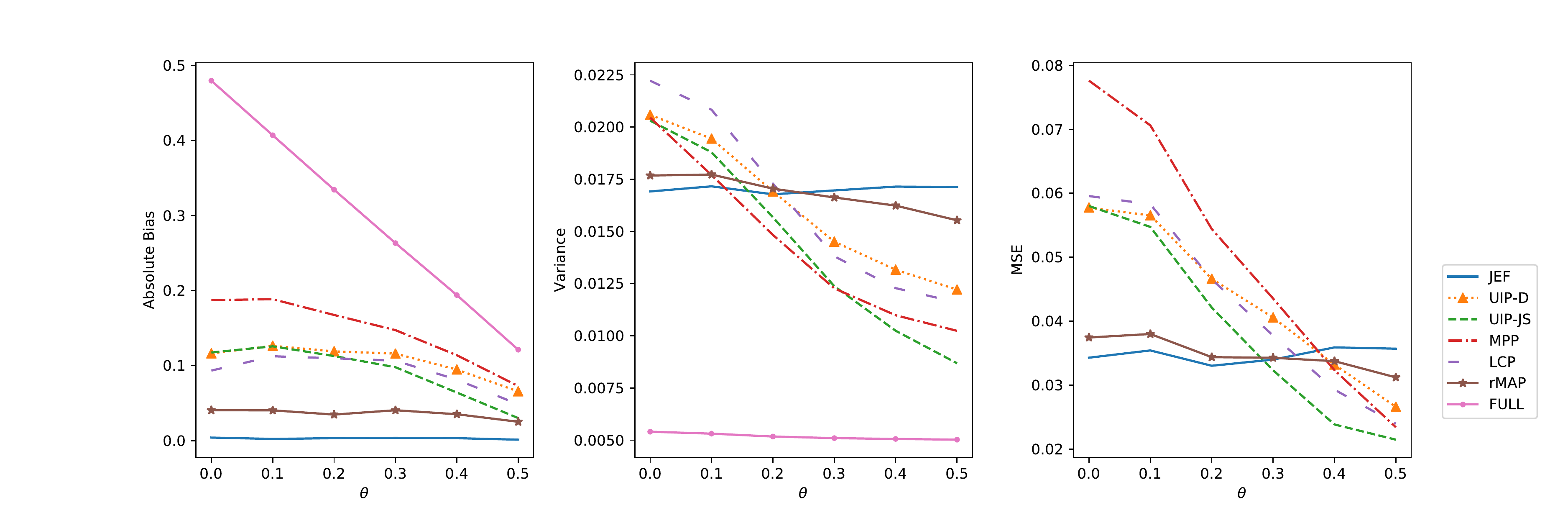}
\end{minipage}%

\begin{minipage}[t]{1\linewidth}
\centering
\includegraphics[scale=0.4]{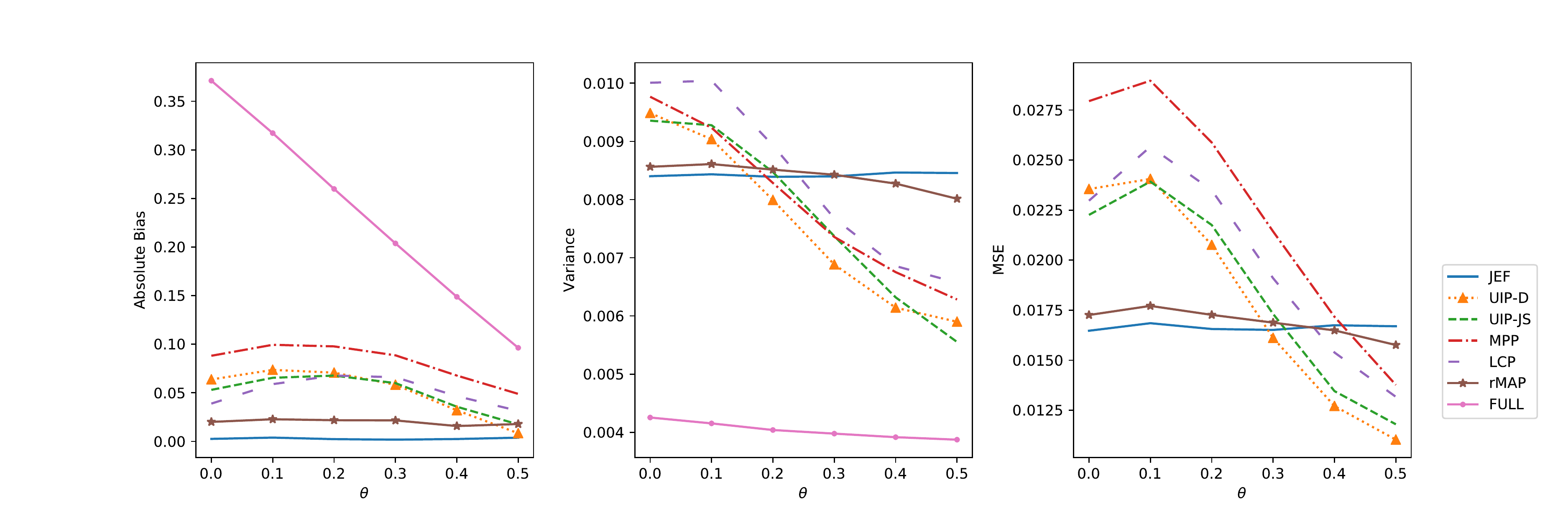}
\end{minipage}%
\caption{
  The absolute bias, variance and mean squared error (MSE) when varying $\theta$ from $0$ to $0.5$
    under Jeffreys prior, UIP-Dirichlet, UIP-JS,  MPP, LCP, rMAP and
    full-borrowing methods
    with sample size $n=60$ (top) and $n=120$ (bottom) for continuous data.
     }
\label{fig:mseNorm}
\end{figure}

To assess the performance of different methods, we show the absolute biases, variances and MSEs
in Figure~\ref{fig:mseNorm} when varying the mean parameter of the current dataset
from $0$ to $0.5$ for $n=60$ and $120$, respectively.
The bias of $\theta$,
defined as $\mathrm{E}\left\{(\theta-\theta_0)|D\right\}$ where $\theta_0$ is the true value,
measures the accuracy for the posterior distribution of $\theta$.
The variance, denoted by $\Var(\theta|D)$, measures the precision of the posterior distribution.
The MSE compromises both the accuracy and precision of the posterior distribution, which
is defined as $\mathrm{E}\left\{(\theta-\theta_0)^2|D\right\}$.
We omit the MSE curve for the full-borrowing method
in order to display other MSE curves better.
We replicate 1000 experiments and take the average for each metric.

All five informative priors show better performances when
the historical datasets are more consistent with the current dataset.
Among them, the rMAP prior yields the most robust results in terms of all three metrics.
However, when the mean parameter of the current dataset is close to the counterparts of the historical datasets,
its variance is only slightly smaller than that under Jefferys' prior.
It illustrates that the rMAP prior tends to be too conservative to borrow enough information
in some cases.
The other four informative priors have a similar trend,
i.e., they all borrow information more aggressively when the historical datasets are coherent with
the current one, yet sacrifice the robustness.
In terms of MSE,
the UIP-JS method is consistently better than the LCP and MPP methods  under both small and large sample sizes.
The UIP-Dirichlet method performs the best under large sample size ($n=120$)
while it has comparable performance with LCP and MPP methods under small sample size ($n=60$).
Larger sample size yields better estimates of the weight parameters
$w_k$'s, which enhance the overall performance of the UIP-Dirichlet method.

\begin{figure}[t]
\centering
\begin{minipage}[t]{1\linewidth}
\centering
\includegraphics[scale=0.4]{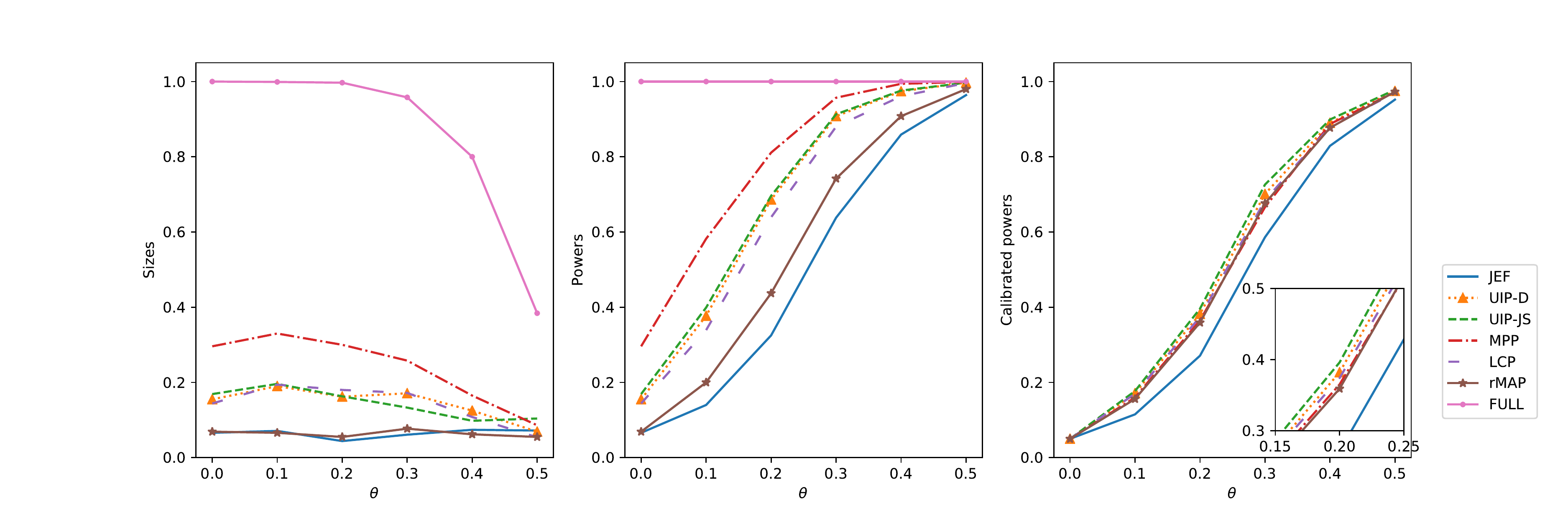}
\end{minipage}%

\begin{minipage}[t]{1\linewidth}
\centering
\includegraphics[scale=0.4]{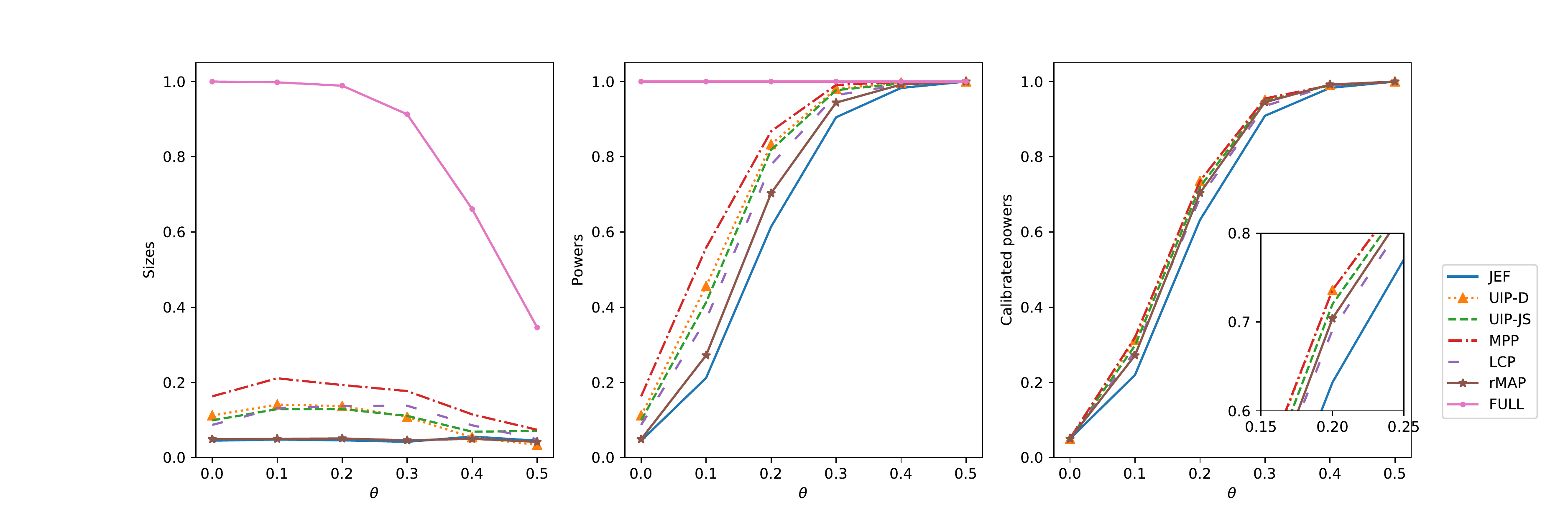}
\end{minipage}%
\caption{Results of
hypothesis testing $H_0:\theta=\theta_0$ vs $H_1: \theta \neq \theta_0$
when varying the true value of the mean parameter from $0$ to $0.5$
for continuous data with sample size $n=60$ (top) and $n=120$ (bottom)
under Jeffreys prior, UIP-Dirichlet, UIP-JS,  MPP, LCP, rMAP and
full-borrowing methods;
left panels: the test size;
middle panels: power; right panels: the calibrated power when controlling the size at $0.05$.
}
\label{fig:testNorm}
\end{figure}

We further conduct hypothesis testing for
$H_0: \theta=\theta_0$ vs $H_1: \theta \neq \theta_0$.
The $95\%$ equal-tailed credible interval (CI) is used as the criterion for the hypothesis testing,
i.e., if the CI does not contain $\theta_0$, we reject $H_0$.
In the left column of Figure~\ref{fig:testNorm},
we present the sizes when varying the
mean parameter for the current dataset from $0$ to $0.5$ with $n=60, 120$
under $1000$ repetitions.
In terms of the type I error,
all five informative priors are robust compared with the-full borrowing method, and
the MPP method has the largest type I error among the five informative priors.
The UIP methods and LCP have comparable sizes, while
the results of the rMAP prior are most close to those of Jeffreys prior.
We show the power
in the middle column of Figure~\ref{fig:testNorm}.
The rMAP prior has the lowest power among the five informative priors
as it is conservative in borrowing information from
historical datasets.
While yielding the largest type I error, the MPP method also has largest power.
The UIP and LCP methods produce comparable results.

To make a fair comparison in power,
we recalibrate the test size of $H_0: \theta=0$ for all seven methods to be 0.05
and present the power curves in the right column of Figure~\ref{fig:testNorm}.
For continuous data, it is impossible to control the size at $0.05$
for the full-borrowing method, which is thus omitted.
After calibration,
there are significant gaps between the informative priors and Jeffreys prior,
revealing that all the informative priors gain information from historical datasets.
The rMAP prior has the overall lowest calibrated power, which
is consistent with the observation that it is conservative in borrowing information.
In fact, in terms of the calibrated power, the UIP methods are consistently better than
the rMAP prior.
Under the small sample size ($n=60$), the UIP-JS method has the largest calibrated power,
while the MPP method leads to the best performance for $n=120$.

The test sizes under the MPP, LCP and UIP methods are significantly inflated compared with
that under Jeffreys prior.
In fact, size inflation is common for informative priors
and a similar phenomenon is observed in the simulations
by \cite{hobbs2011hierarchical, gravestock2019power,banbeta2019modified}.
When incorporating information from historical data to increase the power, it also tends
to inflate the test size.
To solve this issue, we can calibrate the size by enlarging the coverage probability of the
credible interval.
In real data application, the calibration can be implemented by resampling methods, e.g.,
bootstrap or permutation, to reconstruct the null distribution.
In the simulation studies, all the MPP, LCP and UIP methods yield significantly larger power
even after calibration of the size compared with Jeffreys prior
which demonstrates that these informative priors can borrow information from the historical data.
Moreover, not only are informative priors used for frequenstist hypothesis testing,
but they can also help to estimate the parameter of interest.
Figure 3 shows that when the historical datasets do not deviate dramatically from the current dataset,
informative priors can help to decrease the MSE of the parameter estimates
compared with Jeffreys prior.

\section{Application} \label{sec:realdata}
As an illustration, we apply the UIP-Dirichlet and UIP-JS methods to six phase III clinical
studies to investigate the efficacy of memantine in
Alzheimer's disease (AD) \citep{winblad2007memantine,rive2013synthesis}.
All the six trials were double-blind and placebo-controlled.
Among the six trials,
only the trial MRZ-9605 \citep{mrz9605} had a treatment period of $28$ weeks,
while the others had a duration of $24$ weeks. Trials
MEM-MD-02 \citep{memmd02} and MEM-MD-12 \citep{memmd12} took
memantine as an add-on therapy in patients who already received
acetylcholinesterase inhibitors (AChEIs)
and other trials assessed memantine as a monotherapy. Trials
MRZ-9605, MEM-MD-01 \citep{memmd01} and MEM-MD-02 recruited
patients with moderately severe to severe AD,
while trials LU-99679 \citep{lu99679}, MEM-MD-10 \citep{memmd10} and
MEM-MD-12 enrolled patients with mild to moderate AD.
The severity of AD was defined by the scores of the
mini-mental state exam (MMSE).

In our analysis, we regard MEM-MD-12 as the current study
and the remaining trials as historical studies.
The efficacy of the behavioral domain
could be measured at the end of the trial
by the change of the neuropsychiatric inventory (NPI) score from
the baseline, and
a decrease in the NPI score indicated clinical improvement
\citep{cummings1994neuropsychiatric}.
Among the six historical trials, the results of trials LU-99679 and MEM-MD-12
appeared to be closer than others.
For trials LU-99679 and MEM-MD-12, the changes in the NPI scores
indicated that the efficacy of memantine
was inferior to that of placebo in the behavioral domain,
while the rest of the trials demonstrated the opposite results.

Specifically, we analyze the changes of the NPI scores with a linear model,
\bse
Y_i \sim N(\beta_0+\beta_1 X_i, \sigma^2),
\ese
where $Y_i$ is the change of the NPI score for patient $i$, $X_i$ is an
indicator variable taking a value of $1$ if patient $i$ received memantine,
and $0$ otherwise.
Our goal is to determine whether memantine is superior to placebo
in the behavioral domain,
i.e., whether $\beta_1$ is significantly smaller than $0$.
We fit the data of the six trials separately by classical Bayesian linear regression with
noninformative priors for $\beta_0$, $\beta_1$ and $\sigma^2$, i.e.,
$\beta_0 \sim N(0, 10^2)$,
$\beta_1 \sim N(0, 10^2)$ and $\sigma^2 \sim \mathrm{InvGa}(0.01, 0.01)$.

The results in Table~\ref{tab:lm} show that
among all the six studies, MEM-MD-02 is the only trial
with a statistically significant result as its upper bound of the $95\%$
equal-tailed CI
for $\beta_1$ is below $0$.
The estimates of $\beta_1$ in trials LU-99679 and MEM-MD-12 are
positive
and close to each other, and thus we expect that
more information would be borrowed from LU-99679 compared with the
other historical datasets.

We analyze the data from MEM-MD-12 while including the five historical
datasets using the UIP-Dirichlet and UIP-JS methods, in comparison with MPP, LCP and rMAP.
As our main interest focuses on the parameter $\beta_1$, we only impose
the informative priors on $\beta_1$ while for the other parameters we adopt non-informative priors.
To prevent the historical data from overwhelming the current data,
we set $M\sim \mathrm{Uniform}(0, n)$ as the hyper-prior for the total amount parameter $M$ for the UIP methods,
where $n$ is the sample size of the current trial MEM-MD-12.

As shown in Table~\ref{tab:realdatainfo},
all the five informative priors demonstrate the ability of adaptively borrowing
information for $\beta_1$ from historical data
as their $95\%$ CIs of $\beta_1$ are narrower compared with the original CI of $\beta_1$ without any prior information
for MEM-MD-12 in Table~\ref{tab:lm}.
The UIP-Dirichlet, UIP-JS and rMAP methods yield similar results in terms
of $\beta_0$ and $\beta_1$
while the MPP and LCP methods are more analogous with each other.
Among the five informative priors, the $95\%$ CI using the rMAP prior is the widest,
as the rMAP prior is more conservative in borrowing information.
Furthermore, even when we leverage the same non-informative prior for $\beta_0$,
the $95\%$ CIs of $\beta_0$ under UIP and rMAP are narrower than those in Table~\ref{tab:lm},
while for MPP and LCP, the estimates and $95\%$ CIs of $\beta_0$
are essentially unchanged compared with the original results for MEM-MD-12 in Table~\ref{tab:lm}.

For the total amount parameter $M$,
the UIP-Dirichlet and UIP-JS methods lead to comparable results, 137 versus 144,
indicating intermediate borrowing of the historical data
compared with the sample size of the current data $261$.
The ESS of
UIP-Dirichlet is $119$ and that of UIP-JS is $140$, which
justifies the observation in Section~\ref{subsec:ess}, i.e.,
the ESS of UIP is comparable with the corresponding total amount parameter $M$.
Nonetheless, the weight parameters of the two methods are slightly different.
The weight parameters under the UIP-Dirichlet method are
$(0.239, 0.148, 0.217, 0.200, 0.196)$ for the five trials LU-99679,
MEM-MD-01, MEM-MD-02, MEM-MD-10 and MRZ-9605, respectively
while those under the UIP-JS method are $(0.357, 0.155, 0.159, 0.073, 0.256)$.
As expected, both methods assign notably larger weights to the trial LU-99679 compared with
other historical datasets,
while the heterogeneity of the weight parameters under UIP-JS is larger than
that under UIP-Dirichlet.

In summary, under all the five informative priors, although the CIs of $\beta_1$ become narrower,
they still cover $0$.
Thus, in terms of the NPI score,
the efficacy of memantine in the behavioural domain is not
shown to be superior to that of placebo,
which is consistent with the original conclusion in \cite{memmd12}.

The statistics of the NPI scores for both the memantine and placebo
groups of the six trials are presented in the Web Table~1 of the supplementary material,
which also contains more information on the numerical studies.

\section{Discussion}\label{sec:discussion}
We propose an adaptively informative prior using historical data,
which is elicited from an information perspective.
We demonstrate that the UIP framework has many similarities to other commonly used adaptive priors
and yields comparable performances.
The proposed UIP methods are easy to implement for multiple historical datasets,
whose parameters have intuitive interpretations.
The weight parameters $w_k$ can be interpreted as the relative importance
of the historical datasets in comparison with each other.
The amount parameter $M$ reveals the the total information contained in the prior.
For both binary and continuous data, we show that the amount parameter $M$
typically has a comparable value with the prior effective sample size defined by \cite{morita2008determining}.

The UIP methods are useful in the clinical trial field,
as it is not uncommon to find multiple related trials for any ongoing study,
especially, for the control arm of clinical trials.
While we mainly illustrate the UIP methods under the single-arm trial case,
it is also extended to the linear model settings.
In practice, it is typically not easy to obtain the patient-level historical datasets.
An important feature of the UIP framework is that
it does not need the patient-level historical data while
some informative priors (e.g., MPP and LCP methods) need such data.
For example, in a study involving a linear regression model with multiple covariates,
to adopt the UIP-Dirichlet method for the parameter of interest
we only need the estimate of that parameter
and its corresponding confidence interval
which would be commonly reported in publications of the historical study.
However, as the MPP and LCP are derived from the likelihood,
the complete patient-level historical data are required.

\section{Supplementary Material}
\label{sec6}
Web Table 1 shows
the summary statistics of the NPI scores for both the memantine and placebo
groups of six trials in Section~\ref{sec:realdata}.
Web Figure 1 presents
the effective sample sizes of the conditional UIP \citep{morita2008determining} and
marginal UIP \citep{morita2012prior} for binary data.
Web Figure 2 displays the dynamic trend of the amount parameter $M$,
weight parameters $w_k$'s and the absolute weights $Mw_k$'s for binary data.

\section*{Acknowledgments}
The research was supported by a grant No. 17307318 for Guosheng
 Yin from the Research Grants Council of Hong Kong.

	\bibliographystyle{apa.bst}
	\bibliography{sct}%

\begin{thebibliography}{}

\bibitem[\protect\astroncite{Bakchine and Loft}{2008}]{lu99679}
Bakchine, S. and Loft, H. (2008).
\newblock Memantine treatment in patients with mild to moderate alzheimer's
  disease: results of a randomised, double-blind, placebo-controlled 6-month
  study.
\newblock {\em Journal of Alzheimer's Disease}, 13(1):97--107.

\bibitem[\protect\astroncite{Banbeta et~al.}{2019}]{banbeta2019modified}
Banbeta, A., van Rosmalen, J., Dejardin, D., and Lesaffre, E. (2019).
\newblock Modified power prior with multiple historical trials for binary
  endpoints.
\newblock {\em Statistics in medicine}, 38(7):1147--1169.

\bibitem[\protect\astroncite{Borghaei et~al.}{2015}]{borghaei2015nivolumab}
Borghaei, H., Paz-Ares, L., Horn, L., Spigel, D.~R., Steins, M., Ready, N.~E.,
  Chow, L.~Q., Vokes, E.~E., Felip, E., Holgado, E., et~al. (2015).
\newblock Nivolumab versus docetaxel in advanced nonsquamous non--small-cell
  lung cancer.
\newblock {\em New England Journal of Medicine}, 373(17):1627--1639.

\bibitem[\protect\astroncite{Brahmer et~al.}{2015}]{brahmer2015nivolumab}
Brahmer, J., Reckamp, K.~L., Baas, P., Crin{\`o}, L., Eberhardt, W.~E.,
  Poddubskaya, E., Antonia, S., Pluzanski, A., Vokes, E.~E., Holgado, E.,
  et~al. (2015).
\newblock Nivolumab versus docetaxel in advanced squamous-cell non--small-cell
  lung cancer.
\newblock {\em New England Journal of Medicine}, 373(2):123--135.

\bibitem[\protect\astroncite{Cummings
  et~al.}{1994}]{cummings1994neuropsychiatric}
Cummings, J.~L., Mega, M., Gray, K., Rosenberg-Thompson, S., Carusi, D.~A., and
  Gornbein, J. (1994).
\newblock The neuropsychiatric inventory: comprehensive assessment of
  psychopathology in dementia.
\newblock {\em Neurology}, 44(12):2308--2308.

\bibitem[\protect\astroncite{Duan et~al.}{2006}]{duan2006evaluating}
Duan, Y., Ye, K., and Smith, E.~P. (2006).
\newblock Evaluating water quality using power priors to incorporate historical
  information.
\newblock {\em Environmetrics: The Official Journal of the International
  Environmetrics Society}, 17(1):95--106.

\bibitem[\protect\astroncite{Goodfellow
  et~al.}{2014}]{goodfellow2014generative}
Goodfellow, I., Pouget-Abadie, J., Mirza, M., Xu, B., Warde-Farley, D., Ozair,
  S., Courville, A., and Bengio, Y. (2014).
\newblock Generative adversarial nets.
\newblock In {\em Advances in neural information processing systems}, pages
  2672--2680.

\bibitem[\protect\astroncite{Gravestock and Held}{2019}]{gravestock2019power}
Gravestock, I. and Held, L. (2019).
\newblock Power priors based on multiple historical studies for binary
  outcomes.
\newblock {\em Biometrical Journal}, 61(5):1201--1218.

\bibitem[\protect\astroncite{Hobbs et~al.}{2011}]{hobbs2011hierarchical}
Hobbs, B.~P., Carlin, B.~P., Mandrekar, S.~J., and Sargent, D.~J. (2011).
\newblock Hierarchical commensurate and power prior models for adaptive
  incorporation of historical information in clinical trials.
\newblock {\em Biometrics}, 67(3):1047--1056.

\bibitem[\protect\astroncite{Hunter}{2014}]{hunter2014notes}
Hunter, D.~R. (2014).
\newblock Notes for a graduate-level course in asymptotics for statisticians.
\newblock {\em Penn State University, Pennsylvania}.

\bibitem[\protect\astroncite{Ibrahim et~al.}{2000}]{ibrahim2000power}
Ibrahim, J.~G., Chen, M.-H., et~al. (2000).
\newblock Power prior distributions for regression models.
\newblock {\em Statistical Science}, 15(1):46--60.

\bibitem[\protect\astroncite{Itzkovitz et~al.}{2010}]{itzkovitz2010overlapping}
Itzkovitz, S., Hodis, E., and Segal, E. (2010).
\newblock Overlapping codes within protein-coding sequences.
\newblock {\em Genome research}, 20(11):1582--1589.

\bibitem[\protect\astroncite{Kass and Wasserman}{1995}]{kass1995reference}
Kass, R.~E. and Wasserman, L. (1995).
\newblock A reference bayesian test for nested hypotheses and its relationship
  to the schwarz criterion.
\newblock {\em Journal of the American Statistical Association},
  90(431):928--934.

\bibitem[\protect\astroncite{Morita et~al.}{2008}]{morita2008determining}
Morita, S., Thall, P.~F., and M{\"u}ller, P. (2008).
\newblock Determining the effective sample size of a parametric prior.
\newblock {\em Biometrics}, 64(2):595--602.

\bibitem[\protect\astroncite{Morita et~al.}{2012}]{morita2012prior}
Morita, S., Thall, P.~F., and M{\"u}ller, P. (2012).
\newblock Prior effective sample size in conditionally independent hierarchical
  models.
\newblock {\em Bayesian Analysis (Online)}, 7(3).

\bibitem[\protect\astroncite{Neuenschwander
  et~al.}{2009}]{neuenschwander2009note}
Neuenschwander, B., Branson, M., and Spiegelhalter, D.~J. (2009).
\newblock A note on the power prior.
\newblock {\em Statistics in medicine}, 28(28):3562--3566.

\bibitem[\protect\astroncite{Neuenschwander
  et~al.}{2010}]{neuenschwander2010summarizing}
Neuenschwander, B., Capkun-Niggli, G., Branson, M., and Spiegelhalter, D.~J.
  (2010).
\newblock Summarizing historical information on controls in clinical trials.
\newblock {\em Clinical Trials}, 7(1):5--18.

\bibitem[\protect\astroncite{Peskind et~al.}{2006}]{memmd10}
Peskind, E.~R., Potkin, S.~G., Pomara, N., Ott, B.~R., Graham, S.~M., Olin,
  J.~T., McDonald, S., Group, M. M.-M.-.~S., et~al. (2006).
\newblock Memantine treatment in mild to moderate alzheimer disease: a 24-week
  randomized, controlled trial.
\newblock {\em The American Journal of Geriatric Psychiatry}, 14(8):704--715.

\bibitem[\protect\astroncite{Pocock}{1976}]{pocock1976combination}
Pocock, S.~J. (1976).
\newblock The combination of randomized and historical controls in clinical
  trials.
\newblock {\em Journal of chronic diseases}, 29(3):175--188.

\bibitem[\protect\astroncite{Porsteinsson et~al.}{2008}]{memmd12}
Porsteinsson, A.~P., Grossberg, G.~T., Mintzer, J., and Olin, J.~T. (2008).
\newblock Memantine treatment in patients with mild to moderate alzheimer's
  disease already receiving a cholinesterase inhibitor: a randomized,
  double-blind, placebo-controlled trial.
\newblock {\em Current Alzheimer Research}, 5(1):83--89.

\bibitem[\protect\astroncite{Reisberg et~al.}{2003}]{mrz9605}
Reisberg, B., Doody, R., St{\"o}ffler, A., Schmitt, F., Ferris, S., and
  M{\"o}bius, H.~J. (2003).
\newblock Memantine in moderate-to-severe alzheimer's disease.
\newblock {\em New England Journal of Medicine}, 348(14):1333--1341.

\bibitem[\protect\astroncite{Rittmeyer
  et~al.}{2017}]{rittmeyer2017atezolizumab}
Rittmeyer, A., Barlesi, F., Waterkamp, D., Park, K., Ciardiello, F., von Pawel,
  J., Gadgeel, S.~M., Hida, T., Kowalski, D.~M., Dols, M.~C., et~al. (2017).
\newblock Atezolizumab versus docetaxel in patients with previously treated
  non-small-cell lung cancer (oak): a phase 3, open-label, multicentre
  randomised controlled trial.
\newblock {\em The Lancet}, 389(10066):255--265.

\bibitem[\protect\astroncite{Rive et~al.}{2013}]{rive2013synthesis}
Rive, B., Gauthier, S., Costello, S., Marre, C., and Fran{\c{c}}ois, C. (2013).
\newblock Synthesis and comparison of the meta-analyses evaluating the efficacy
  of memantine in moderate to severe stages of alzheimer’s disease.
\newblock {\em CNS Drugs}, 27(7):573--582.

\bibitem[\protect\astroncite{R{\"o}ver and Friede}{2020a}]{rover2020bounds}
R{\"o}ver, C. and Friede, T. (2020a).
\newblock Bounds for the weight of external data in shrinkage estimation.
\newblock {\em arXiv preprint arXiv:2004.02525}.

\bibitem[\protect\astroncite{R{\"o}ver and
  Friede}{2020b}]{rover2020dynamically}
R{\"o}ver, C. and Friede, T. (2020b).
\newblock Dynamically borrowing strength from another study through shrinkage
  estimation.
\newblock {\em Statistical Methods in Medical Research}, 29(1):293--308.

\bibitem[\protect\astroncite{Schmidli et~al.}{2014}]{schmidli2014robust}
Schmidli, H., Gsteiger, S., Roychoudhury, S., O'Hagan, A., Spiegelhalter, D.,
  and Neuenschwander, B. (2014).
\newblock Robust meta-analytic-predictive priors in clinical trials with
  historical control information.
\newblock {\em Biometrics}, 70(4):1023--1032.

\bibitem[\protect\astroncite{Sims et~al.}{2009}]{sims2009alignment}
Sims, G.~E., Jun, S.-R., Wu, G.~A., and Kim, S.-H. (2009).
\newblock Alignment-free genome comparison with feature frequency profiles
  (ffp) and optimal resolutions.
\newblock {\em Proceedings of the National Academy of Sciences},
  106(8):2677--2682.

\bibitem[\protect\astroncite{Tariot et~al.}{2004}]{memmd02}
Tariot, P.~N., Farlow, M.~R., Grossberg, G.~T., Graham, S.~M., McDonald, S.,
  Gergel, I., Group, M.~S., et~al. (2004).
\newblock Memantine treatment in patients with moderate to severe alzheimer
  disease already receiving donepezil: a randomized controlled trial.
\newblock {\em Journal of the American Medical Association}, 291(3):317--324.

\bibitem[\protect\astroncite{van Dyck et~al.}{2007}]{memmd01}
van Dyck, C.~H., Tariot, P.~N., Meyers, B., Resnick, E.~M., Group, M.
  M.-M.-.~S., et~al. (2007).
\newblock A 24-week randomized, controlled trial of memantine in patients with
  moderate-to-severe alzheimer disease.
\newblock {\em Alzheimer Disease \& Associated Disorders}, 21(2):136--143.

\bibitem[\protect\astroncite{Weber et~al.}{2019}]{weber2019applying}
Weber, S., Li, Y., Seaman, J., Kakizume, T., and Schmidli, H. (2019).
\newblock Applying meta-analytic predictive priors with the r bayesian evidence
  synthesis tools.
\newblock {\em arXiv preprint arXiv:1907.00603}.

\bibitem[\protect\astroncite{Winblad et~al.}{2007}]{winblad2007memantine}
Winblad, B., Jones, R.~W., Wirth, Y., St{\"o}ffler, A., and M{\"o}bius, H.~J.
  (2007).
\newblock Memantine in moderate to severe alzheimer’s disease: a
  meta-analysis of randomised clinical trials.
\newblock {\em Dementia and Geriatric Cognitive Disorders}, 24(1):20--27.

\bibitem[\protect\astroncite{Wu et~al.}{2019}]{wu2019nivolumab}
Wu, Y.-L., Lu, S., Cheng, Y., Zhou, C., Wang, J., Mok, T., Zhang, L., Tu,
  H.-Y., Wu, L., Feng, J., et~al. (2019).
\newblock Nivolumab versus docetaxel in a predominantly chinese patient
  population with previously treated advanced nsclc: Checkmate 078 randomized
  phase iii clinical trial.
\newblock {\em Journal of Thoracic Oncology}, 14(5):867--875.

\end{thebibliography}


\begin{thebibliography}{}

\bibitem[\protect\astroncite{Morita et~al.}{2008}]{morita2008determining}
Morita, S., Thall, P.~F., and M{\"u}ller, P. (2008).
\newblock Determining the effective sample size of a parametric prior.
\newblock {\em Biometrics}, 64(2):595--602.

\bibitem[\protect\astroncite{Morita et~al.}{2012}]{morita2012prior}
Morita, S., Thall, P.~F., and M{\"u}ller, P. (2012).
\newblock Prior effective sample size in conditionally independent hierarchical
  models.
\newblock {\em Bayesian Analysis (Online)}, 7(3).

\end{thebibliography}

\newpage

\clearpage
\newpage

\begin{table}[h!]
    \centering
    \caption{The results of Bayesian linear regression fitted separately to
    each of the six trials and the corresponding weight assigned to each dataset
    when using UIP-Dirichlet and UIP-JS respectively.}
    \begin{tabular}{lcccccc}
    \toprule
    & \multicolumn{2}{c}{$\beta_0$} & \multicolumn{2}{c}{$\beta_1$} & {UIP-Dirichlet}& {UIP-JS}\\
    \cmidrule(r){2-5}
    Trials       & Estimate & $95\%$ CI & Estimate & $95\%$ CI&Weight &Weight\\
    \midrule
    MEM-MD-12 &  0.865 &  (-1.065, 2.838) &   0.093 &  (-2.553, 2.801)& 1 & 1 \\
    LU-99679  & -2.173 &  (-4.595, 0.349) &   1.819 &  (-1.185, 4.744)&0.239& 0.357\\
    MEM-MD-01 &  0.479 &  (-2.012, 2.974) &  -2.568 &  (-5.994, 0.950)&0.148& 0.155\\
    MEM-MD-02 &  2.701 &   (0.811, 4.650) &  -3.412 & (-6.094, -0.745)&0.217& 0.159\\
    MEM-MD-10 &  2.796 &   (0.233, 5.318) &  -2.017 &  (-5.595, 1.725)&0.200& 0.073\\
    MRZ-9605  &  2.704 &  (-0.541, 6.045) &  -2.541 &  (-7.058, 1.968)&0.196& 0.256\\
    \bottomrule
    \multicolumn{5}{l}{\footnotesize CI: credible interval.}
    \end{tabular}
    \label{tab:lm}
\end{table}

\begin{table}[h!]
    \centering
    \caption{The results using the UIP-Dirichlet, UIP-JS, MPP, LCP and rMAP methods for
    the current study MEM-MD-12 by borrowing information
    from five historical trials LU-99679, MEM-MD-01, MEM-MD-02, MEM-MD-10 and MRZ-9605. }
    \begin{tabular}{lcccccc}
    \toprule
         \multicolumn{2}{l}{Parameters} & UIP-Dirichlet & UIP-JS & MPP & LCP & rMAP  \\
         \midrule
    $\beta_0$ & Estimate  & 1.232         & 1.128         & 0.856         & 0.849  & 1.166 \\
              & $95\%$ CI &(-0.560, 2.974)&(-0.571, 2.851)&(-1.127, 2.809)&(-1.104, 2.780)&(-0.705, 2.942)\\
    $\beta_1$ & Estimate  &-0.626         & -0.400        &-0.837         &-0.746 &-0.476\\
              & $95\%$ CI &(-2.783, 1.643)&(-2.350, 1.639)&(-3.132, 1.502)&(-3.061, 1.639)&(-2.740, 2.020)\\
    \bottomrule
    \end{tabular}
    \label{tab:realdatainfo}
\end{table}
\end{document}